\documentclass[aps,pra,twocolumn,showpacs,amsmath,amssymb,superscriptaddress]{revtex4-1}

\usepackage{graphicx}
\usepackage{dcolumn}
\usepackage{bm}

\begin{document}

\title{Quantum-classical transition of correlations of two coupled cavities}

\author{Tony E. Lee}
\affiliation
{ITAMP, Harvard-Smithsonian Center for Astrophysics, Cambridge, MA 02138, USA}
\affiliation{Department of Physics, California Institute of Technology, Pasadena, California 91125, USA}
\author{M. C. Cross}
\affiliation{Department of Physics, California Institute of Technology, Pasadena, California 91125, USA}

\date{\today}

\begin{abstract}
We study the difference between quantum and classical behavior in a pair of nonidentical cavities with second-harmonic generation. In the classical limit, each cavity has a limit-cycle solution, in which the photon number oscillates periodically in time. Coupling between the cavities leads to synchronization of the oscillations and classical correlations between the cavities. In the quantum limit, there are quantum correlations due to entanglement. The quantum correlations persist even when the cavities are far off resonance with each other, in stark contrast with the classical case. We also find that the quantum and classical limits are connected by an intermediate regime of almost no correlations. Our results can be extended to a wide variety of quantum models.
\end{abstract}

\maketitle

\section{Introduction}
A central theme in quantum optics is to distinguish between quantum and classical behavior \cite{walls06}. For a single optical cavity, this can be done by measuring photon correlations \cite{birnbaum05} or transmission spectra \cite{fink10}. A fundamental question is then: when there are two cavities, how does their collective behavior differ in the quantum and classical regimes? Since optical cavities are inherently dissipative due to photon decay, it is natural to look at photon correlations between the cavities \cite{liew10,nissen12}. Thus, one would like to see how entanglement (a uniquely quantum feature) affects the photon correlations.

However, the mere presence of photon correlations between the cavities is \emph{not} a uniquely quantum effect, since a classical system can also have photon correlations. The reason for classical correlations is as follows. It turns out that many quantum optical models exhibit limit cycles in the classical limit, e.g., when there are many photons in the cavity \cite{drummond80,hu90,armen06,kippenberg05,marquardt06,ludwig08,lee11b,qian12,ono04,rodrigues07,lee13}. A limit cycle means that the photon number oscillates in steady state. Then when the cavities are coupled, the limit cycles synchronize with each other \cite{pikovsky01,strogatz03,acebron05}, leading to bunching or antibunching between the cavities. This is a purely classical effect, since the classical model assumes no entanglement.

Since experiments are reaching the point of observing collective quantum effects \cite{underwood12,majumdar12,zhang12,carusotto13}, it is important now to distinguish between quantum and classical photon correlations. It is also of fundamental interest to study how classical correlations from synchronization turn into quantum correlations from entanglement as the system becomes more quantum. If one could experimentally measure the complete density matrix of the coupled system, one could apply various entanglement measures to distinguish between quantum and classical correlations \cite{kaszlikowski08,modi10,gallego10}; however, it is experimentally much easier to measure photon correlations, which is the approach we take.


In this paper, we consider two coupled cavities, each with second-harmonic generation. In the classical limit, each cavity has a limit-cycle solution \cite{drummond80}, and coupling between the cavities causes the oscillations to synchronize. We make the system more quantum by decreasing the number of photons in the cavities. We find that quantum correlations can be distinguished from classical correlations by detuning the cavities from each other. In contrast to the classical case, quantum correlations remain even when the cavities are far detuned from each other, and they can even be stronger than when they are identical. We also elucidate the nature of the quantum-classical transition: the quantum and classical limits are connected by an intermediate regime, in which both types of correlations are washed out by quantum noise.




Although we focus on second-harmonic generation, the physics we invoke is quite general, and our results can be extended to the many other quantum models known to have limit cycles: Jaynes-Cummings cavities \cite{hu90,armen06}, optomechanics \cite{kippenberg05,marquardt06,ludwig08}, Rydberg atoms \cite{lee11b,qian12}, quantum dots \cite{ono04}, single-electron transistors \cite{rodrigues07}, and trapped ions \cite{lee13}.

\begin{figure}[b]
\centering
\includegraphics[width=2.5 in,trim=2in 3.5in 2in 0.5in,clip]{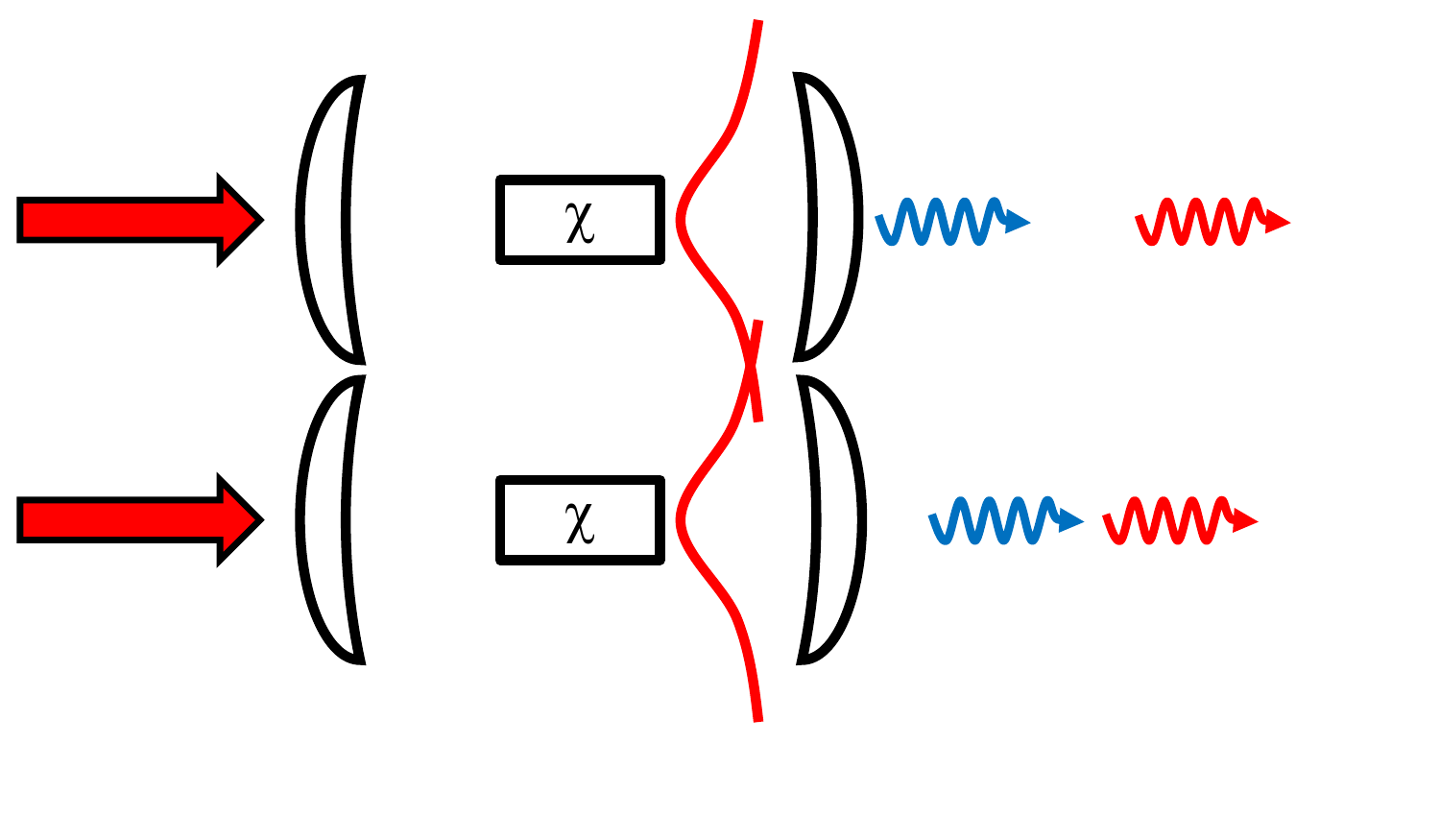}
\caption{\label{fig:cartoon}Two cavities with nonlinear crystals inside are laser-driven and dissipate photons. They are coupled to each other due to overlap of their photonic wave functions.}
\end{figure}


\section{Model}
In second-harmonic generation, a nonlinear crystal within an optical cavity converts light at a fundamental frequency into light at twice the frequency \cite{franken61}. We consider the case of two cavities, each with a nonlinear crystal inside. The cavities are coupled due to overlap of their photonic wave functions (Fig.~\ref{fig:cartoon}). 

We first describe the quantum model and then the classical model. Let the two cavities be denoted $a$ and $b$. Each cavity has two modes: $a_1$ and $a_2$ are the annihilation operators for the fundamental and second-harmonic modes of the first cavity, while $b_1$ and $b_2$ are the corresponding operators for the second cavity. An external laser drives the fundamental mode of both cavities, and the nonlinear crystals produce light at the second harmonic. In the interaction picture and rotating-wave approximation, the Hamiltonian is ($\hbar=1$)
\begin{eqnarray}
H&=&iE(a_1^\dagger-a_1+b_1^\dagger-b_1)\nonumber\\
&&+i\frac{\chi}{2}(a_1^{\dagger2}a_2-a_1^2a_2^\dagger+b_1^{\dagger2}b_2-b_1^2b_2^\dagger)\nonumber\\
&&+\Delta_1^a a_1^\dagger a_1+ \Delta_1^b b_1^\dagger b_1 + \Delta_2(a_2^\dagger a_2+b_2^\dagger b_2)\nonumber\\
&&+V_1(a_1^\dagger b_1+a_1b_1^\dagger)+V_2(a_2^\dagger b_2+a_2b_2^\dagger),\label{eq:H}
\end{eqnarray}
where $E$ is the laser drive and $\chi$ is the second-order susceptibility of the crystal. $\Delta_1^a$ and $\Delta_1^b$ are the detunings of the fundamental modes of the two cavities  from the laser. We allow $\Delta_1^a$ and $\Delta_1^b$ to differ from each other, in order to see how their disparity affects the synchronization. $\Delta_2$ is the frequency detuning of the second harmonics from the laser. To be precise, $\Delta_1^a\equiv\omega_o^a-\omega_{\ell}^a$, where $\omega_o^a$ is the fundamental frequency of cavity $a$, and $\omega_{\ell}^a$ is the frequency of the laser that drives it. Similarly, $\Delta_1^b\equiv\omega_o^b-\omega_{\ell}^b$. $V_1$ ($V_2$) is the coupling between the fundamental (second-harmonic) modes. The term with $\chi$ means that two photons at the fundamental frequency are converted into one photon at the second harmonic; the reverse process is also allowed. 



Photons leak out of the cavities with rates $\kappa_1$ and $\kappa_2$ for the fundamental and second harmonic, respectively. This open quantum system is described by a Lindblad master equation for the density matrix $\rho$:
\begin{eqnarray}
\dot{\rho}&=&-i[H,\rho]+\sum_{i=1,2}\kappa_i(2a_i\rho a_i^\dagger-a_i^\dagger a_i\rho-\rho a_i^\dagger a_i)\nonumber\\
&&\quad\quad\quad\quad+\sum_{i=1,2}\kappa_i(2b_i\rho b_i^\dagger-b_i^\dagger b_i\rho-\rho b_i^\dagger b_i),\label{eq:master}
\end{eqnarray}
which is linear in $\rho$ and has a unique steady state \cite{schirmer10}.

In the classical approximation to Eq.~\eqref{eq:master}, one 
replaces the operators $a_1,a_2,b_1,b_2$ with complex numbers that denote coherent states $\alpha_1,\alpha_2,\beta_1,\beta_2$. This leads to classical equations of motion that are nonlinear:
\begin{eqnarray}
\dot{\alpha}_1&=&E-(\kappa_1+i\Delta_1^a)\alpha_1+\chi\alpha_1^*\alpha_2-iV_1\beta_1,\label{eq:alpha1}\\
\dot{\alpha}_2&=&\quad\,-(\kappa_2+i\Delta_2)\alpha_2-\frac{\chi}{2}\alpha_1^2-iV_2\beta_2,\label{eq:alpha2}\\
\dot{\beta}_1&=&E-(\kappa_1+i\Delta_1^b)\beta_1+\chi\beta_1^*\beta_2-iV_1\alpha_1,\label{eq:beta1}\\
\dot{\beta}_2&=&\quad\,-(\kappa_2+i\Delta_2)\beta_2-\frac{\chi}{2}\beta_1^2-iV_2\alpha_2.\label{eq:beta2}
\end{eqnarray}
The average number of photons in mode $a_1$ is $\langle a_1^\dagger a_1\rangle=|\alpha_1|^2$ and similarly for other modes. The classical approximation is an accurate description of the quantum model when there are many photons in each mode \cite{zheng95}. 
(This occurs when the laser drive is much stronger than the dissipation, since the photon number is determined by the balance of driving and dissipation.) Intuitively, this is because when a mode is highly populated in steady state, it continuously emits photons, so an individual photon emission has negligible effect.

It is insightful to rewrite Eqs.~\eqref{eq:alpha1}--\eqref{eq:beta2} using scaled variables ($\tilde{\alpha}_i=c\alpha_i$, $\tilde{\beta}_i=c\beta_i$) and scaled parameters ($\tilde{E}=cE$, $\tilde{\chi}=\chi/c$), where $c$ is an arbitrary number. It turns out that the dynamics is independent of $c$, up to a scaling of $\alpha_i$ and $\beta_i$. 
This provides a controlled way of following the classical-to-quantum transition \cite{zheng95}. We will solve the quantum model while decreasing $E$ and increasing $\chi$, keeping $E\chi$ fixed, so that the photon numbers decrease. Since the classical dynamics remains the same in this procedure, \emph{any change in behavior must be due to quantum effects}. The quantum limit corresponds to small $E$ and large $\chi$.

Note that all the parameters (like $\chi$ and $E$) are given in units of $\kappa_1$, since one can scale out $\kappa_1$ from Eq.~\eqref{eq:master} by redefining time. In practice, the absolute value of $\chi$ is fixed, so to see the classical-to-quantum transition, one would decrease $\kappa_1$.

To measure the photon correlations between the various modes, we calculate:
\begin{eqnarray}
g_2(a_1,a_2)=\frac{\langle a_1^\dagger a_1a_2^\dagger a_2\rangle}{\langle a_1^\dagger a_1\rangle\langle a_2^\dagger a_2\rangle}, \quad
g_2(a_i,b_i)=\frac{\langle a_i^\dagger a_ib_i^\dagger b_i\rangle}{\langle a_i^\dagger a_i\rangle\langle b_i^\dagger b_i\rangle}.\nonumber \label{eq:g2a1b1}
\end{eqnarray}
When $g_2>1$, the two modes are positively correlated and tend to emit photons simultaneously (bunching). When $g_2<1$, they are negatively correlated and tend not to emit simultaneously (antibunching). When $g_2=1$, there are no correlations. We will use $g_2$ as an order parameter, and it is easy to measure experimentally.
In the quantum model, expectation values are taken with respect to the steady state $\rho$. In the classical model, expectation values are averages over time.

\section{One cavity, classical model}
\begin{figure}
\centering
\includegraphics[width=3.4 in,trim=1in 3.8in 1in 3.8in,clip]{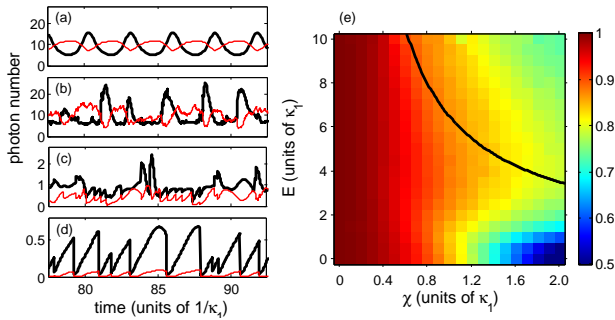}
\caption{\label{fig:N1_trajectories}Classical and quantum trajectories for one cavity, showing photon numbers of mode $a_1$ (thick, black line) and mode $a_2$ (thin, red line) over time. (a) Limit-cycle solution of the classical model for $E=8\kappa_1,\chi=0.8\kappa_1$. (b)--(d) Quantum trajectories for the same system, but as it becomes more quantum: (b) $E=8\kappa_1,\chi=0.8\kappa_1$; (c) $E=2\kappa_1,\chi=3.2\kappa_1$; (d) $E=\kappa_1,\chi=6.4\kappa_1$. In (d), the modes are antibunched with each other due to quantum correlations, despite appearing to be positively correlated in the plot. (e) Correlation $g_2(a_1,a_2)$ for one cavity, using color scale on right. The black line indicates the location of the Hopf bifurcation $E_c$. All plots use $\kappa_2=0.5\kappa_1,\Delta_1=0.5\kappa_1,\Delta_2=\kappa_1$.}
\end{figure}

Here we review the results for one classical cavity, i.e., Eqs.~\eqref{eq:alpha1} and \eqref{eq:alpha2} with $V_1=V_2=0$. When $E$ is at a critical value $E_c$, a Hopf bifurcation occurs  \cite{drummond80}. When $E<E_c$, the system has a stable fixed-point solution. When $E>E_c$, the fixed point is unstable, and a limit cycle appears. This means that the photon numbers $|\alpha_1|^2$ and $|\alpha_2|^2$ oscillate periodically in time [Fig.~\ref{fig:N1_trajectories}(a)]. 
$E_c$ is a function of $\chi$, $\kappa_i$, $\Delta_i$. 

The existence of the limit cycle can be intuitively understood as follows. When $E$ is small, $\alpha_1$ and $\alpha_2$ are small, so the nonlinear terms proportional to $\chi$ in Eqs.~\eqref{eq:alpha1} and \eqref{eq:alpha2} have negligible effect. But for sufficiently large $E$, $\alpha_1$ and $\alpha_2$ are large, and the $\chi$ terms dominate. The effect of the $\chi$ terms is to exchange energy back and forth between the two modes. This exchange is seen in Fig.~\ref{fig:N1_trajectories}(a), where $|\alpha_1|^2$ and $|\alpha_2|^2$ are roughly anti-phase with each other.
The existence of the limit cycle is indicated by $g_2(a_1,a_2)$. When $E<E_c$, there are no correlations: $g_2(a_1,a_2)=1$. When $E>E_c$, the modes are negatively correlated: $g_2(a_1,a_2)<1$.
These correlations are completely classical.



\section{One cavity, quantum model}
As one cavity becomes more quantum, the limit cycle becomes noisier \cite{savage88}, as seen in quantum trajectory simulations [Figs. \ref{fig:N1_trajectories}(b)--(d)] \cite{dalibard92,dum92}. In the extreme quantum limit, when there is much less than one photon in each mode, the limit cycle is not visually identifiable at all. In this limit, there are strong quantum correlations between $a_1$ and $a_2$, found by perturbatively solving for the steady state of Eq.~\eqref{eq:master}:
\begin{eqnarray}
g_2(a_1,a_2)=\frac{1}{1+\frac{\frac{1}{4}\chi^4+\chi^2[-\Delta_1(\Delta_1+\Delta_2)+\kappa_1(\kappa_1+\kappa_2)]}{(\Delta_1^2+\kappa_1^2)[(\Delta_1+\Delta_2)^2+(\kappa_1+\kappa_2)^2]}}+O(E^2). \nonumber \label{eq:g2a1a2_quantum}
\end{eqnarray}
When $E\rightarrow0$ and $\chi\rightarrow\infty$, $g_2(a_1,a_2)=0$ (strong antibunching). 
Figure \ref{fig:N1_trajectories}(e) shows the quantum-classical transition. In the classical limit, $g_2(a_1,a_2)=1$ when $E<E_c$ and $g_2(a_1,a_2)<1$ when $E>E_c$. As the system becomes more quantum, the transition at $E_c$ smoothes out, and the region of antibunching expands into $E<E_c$. Antibunching in the classical limit gradually develops into strong antibunching in the quantum limit. Thus, the quantum remnant of the limit cycle retains the antibunching signature.

\begin{figure}[t]
\centering
\includegraphics[width=3.5 in,trim=0in 3in 0in 0in,clip]{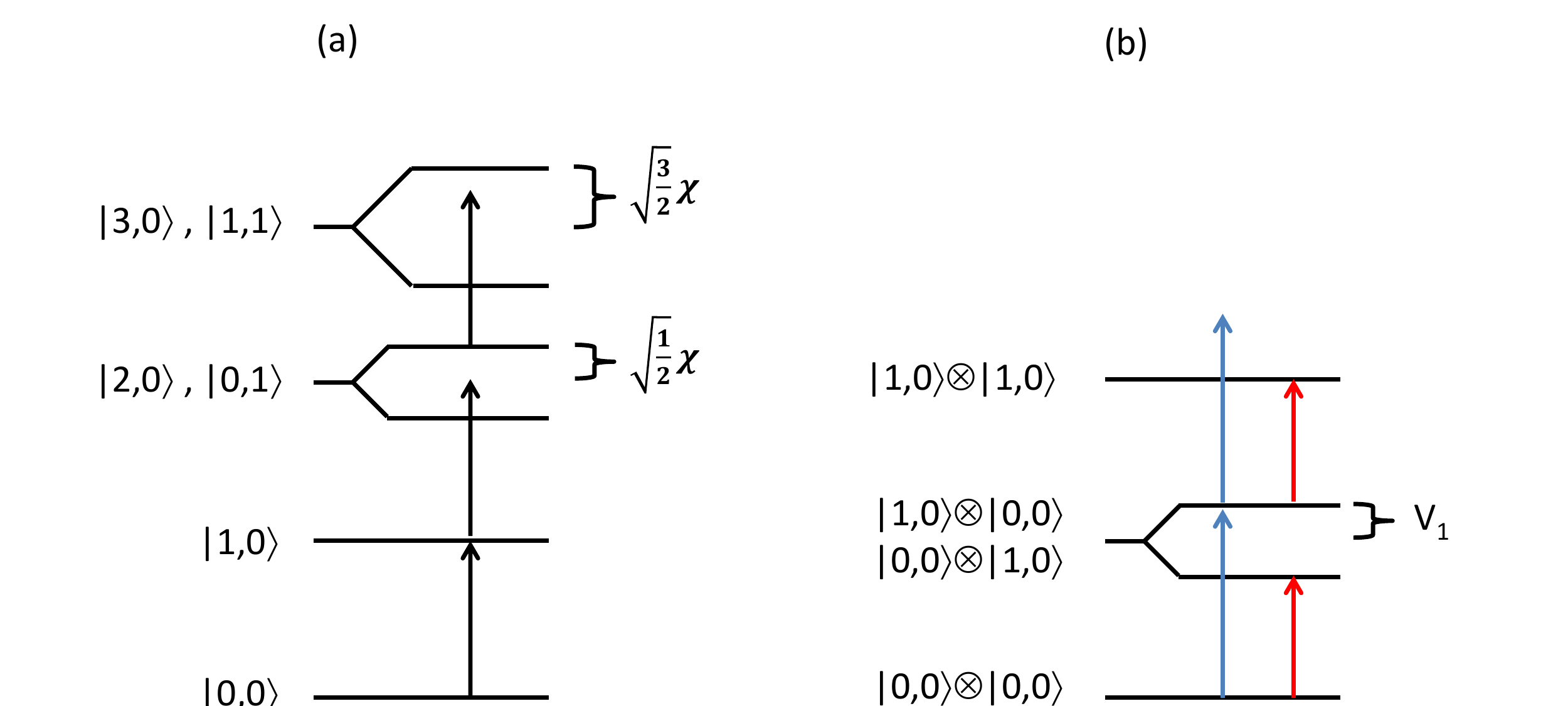}
\caption{\label{fig:N1_levels}(a) Eigenstates for one cavity. (b) Eigenstates for two identical cavities. When $\Delta_1<0$, there is antibunching between $a_1$ and $b_1$ (blue arrows). When $\Delta_1>0$, there is antibunching (red arrows).}
\end{figure}

The physical reason for antibunching can be understood by examining the eigenstates of the Hamiltonian in the absence of the laser drive: 
\begin{eqnarray}
H_1&=&i\frac{\chi}{2}(a_1^{\dagger2}a_2-a_1^2a_2^\dagger).
\end{eqnarray}
One should think of the laser as exciting the eigenstates of this Hamiltonian, which are shown in Fig.~\ref{fig:N1_levels}(a). Let the basis be $|m_1,m_2\rangle$, where $m_1$ is the Fock state of the fundamental mode, and $m_2$ is the Fock state of the second harmonic. The degeneracy of $|2,0\rangle$ and $|0,1\rangle$ is lifted by $\chi$, resulting in two eigenstates that are each shifted by $\sqrt{\frac{1}{2}}\chi$. The eigenstates are $|2,0\rangle\pm i|0,1\rangle$. Similarly, the degeneracy of $|3,0\rangle$ and $|1,1\rangle$ is lifted by $\chi$, resulting in two eigenstates that are each shifted by $\sqrt{\frac{3}{2}}\chi$.

Now, whether $a_1$ and $a_2$ are bunched or antibunched in the quantum limit depends on the relative population of $|1,1\rangle$, since that determines whether the two modes can emit at the same time. But in the limit $\chi\rightarrow\infty$, the level shifts of the eigenstates are infinite. Thus, the laser is unable to excite any eigenstates above $|1,0\rangle$. As a result, $g_2(a_1,a_2)=0$.

\section{Two cavities, classical model}
Now we study the synchronization of two cavities in the classical limit. Following convention, we assume that $V_i$ is small, so that the limit cycle of each cavity retains its identity \cite{pikovsky01}. Note that synchronization is universal in the sense than any system with coupled limit cycles will exhibit synchronization \cite{pikovsky01}.

\begin{figure}[t]
\centering
\includegraphics[width=3.4 in,trim=1in 4.2in 1in 4.2in,clip]{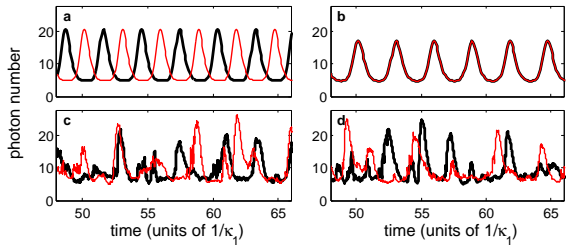}
\caption{\label{fig:N2_sync}Classical and quantum trajectories for two identical cavities, showing photon numbers of mode $a_1$ (thick, black line) and mode $b_1$ (thin, red line) over time. (a) Anti-phase synchrony for $\Delta_1^a=\Delta_1^b=0.5\kappa_1,\Delta_2=\kappa_1$ in the classical model. (b) In-phase synchrony for $\Delta_1^a=\Delta_1^b=-0.5\kappa_1,\Delta_2=-\kappa_1$ in the classical model. (c,d) Quantum trajectories for the same parameters as (a) and (b), respectively. All plots use $E=8\kappa_1,\chi=0.8\kappa_1,\kappa_2=0.5\kappa_1,V_1=0.2\kappa_1,V_2=0$.}
\end{figure}

\begin{figure}[t]
\centering
\includegraphics[width=3.4 in,trim=1in 4.2in 1in 4.2in,clip]{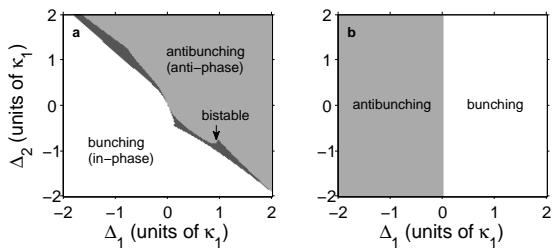}
\caption{\label{fig:N2_phasediagram}Phase diagram for two identical cavities as a function of $\Delta_1\equiv\Delta_1^a=\Delta_1^b$ and $\Delta_2$. (a) Classical limit for $\chi=0.8\kappa_1,\kappa_2=0.5\kappa_1,V_1=0.2\kappa_1,V_2=0$. For each value of $\Delta_1$ and $\Delta_2$, $E$ is set to $E_c+0.5\kappa_1$. (b) Quantum limit for the same parameters, except with $E\rightarrow0$ and $\chi\rightarrow\infty$.}
\end{figure}

When the cavities are identical ($\Delta_1^a=\Delta_1^b$), the steady state will be either anti-phase [Fig.~\ref{fig:N2_sync}(a)] or in-phase oscillation [Fig.~\ref{fig:N2_sync}(b)] \cite{aronson90,lee11a}. The first solution means that $g_2(a_1,b_1),g_2(a_2,b_2)<1$, while the second means that $g_2(a_1,b_1),g_2(a_2,b_2)>1$. The phase diagram is shown in Fig.~\ref{fig:N2_phasediagram}(a) \footnote{This is shown by reducing the equations to the normal form of a Hopf bifurcation \cite{nayfeh93}, and then adiabatically eliminating the amplitudes to obtain a phase equation \cite{pikovsky01}}. Note that these correlations are completely classical.


\begin{figure}
\centering
\includegraphics[width=3.4 in,trim=1in 4.5in 1in 4.5in,clip]{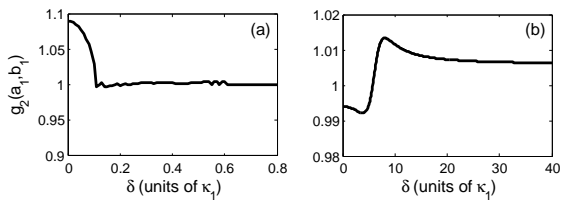}
\caption{\label{fig:N2_nonidentical}Correlation $g_2(a_1,b_1)$ for two nonidentical cavities vs. the difference $\delta\equiv\Delta_1^a-\Delta_1^b$ for a fixed average $\Delta_1\equiv(\Delta_1^a+\Delta_1^b)/2$. (a) Classical limit for $E=10.5\kappa_1,\chi=\kappa_1,\kappa_2=0.5\kappa_1,\Delta_1=-3\kappa_1,\Delta_2=V_2=0,V_1=0.01\kappa_1$. (b) Quantum limit for the same parameters, except with $E\rightarrow 0,\chi\rightarrow\infty$.}
\end{figure}

When the cavities are nonidentical ($\Delta_1^a\neq\Delta_1^b$), the limit cycles of the two cavities have different intrinsic frequencies. When the difference in intrinsic frequencies is small relative to the coupling, the limit cycles phase lock with each other and oscillate with the same frequency [$g_2(a_1,b_1)\neq 1$ since there are still correlations]. When the difference is large relative to the coupling, the limit cycles do not lock and they continue to oscillate with different frequencies [$g_2(a_1,b_1)\approx 1$ since the cavities are not correlated]. Figure \ref{fig:N2_nonidentical}(a) shows $g_2(a_1,b_1)$ as the difference between $\Delta_1^a$ and $\Delta_1^b$ grows, for a fixed coupling strength. The synchronization transition occurs suddenly at a critical difference.

Suppose one adds noise to a synchronized system. When the noise is weak, the limit cycles experience occasional phase slips relative to each other, but they remain mostly phase-locked with each other \cite{pikovsky01}. As the noise level increases, phase slips occur more often. Since $g_2(a_1,b_1)$ measures the correlation between the cavities, as the noise level increases, $g_2(a_1,b_1)$ gradually approaches 1 from its no-noise value. [Numerical experiments with Eqs.~\eqref{eq:alpha1}--\eqref{eq:beta2} show that this is true.] This reflects the fact that noise inhibits synchronization.




\section{Two cavities, quantum model}
We first consider two identical cavities ($\Delta_1^a=\Delta_1^b$) in the quantum limit. We want to see what happens to these correlations in the quantum limit. Since adding increasing amounts of classical noise causes $g_2(a_1,b_1)$ to approach 1, one would expect $g_2(a_1,b_1)=1$ in the quantum limit due to substantial quantum noise.


\begin{figure}
\centering
\includegraphics[width=3.4 in,trim=1in 4.5in 1in 4.5in,clip]{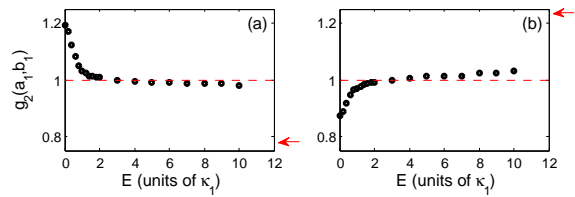}
\caption{\label{fig:N2_g2}Quantum-classical transition for two identical cavities. For each value of $E$, $\chi$ is set so that $E\chi=6.4\kappa_1^2$. (a) $\Delta_1^a=\Delta_1^b=0.5\kappa_1$, $\Delta_2=\kappa_1$. (b) $\Delta_1^a=\Delta_1^b=-0.5\kappa_1$, $\Delta_2=-\kappa_1$. Red arrows point to the classical values of $g_2(a_1,b_1)$. Dashed lines mark $g_2(a_1,b_1)=1$. Other parameters are $\kappa_2=0.5\kappa_1$, $V_1=0.2\kappa_1$, $V_2=0$. The statistical uncertainty for each point is smaller than the marker size.}
\end{figure}

Figure \ref{fig:N2_g2} shows the quantum-classical transition. Indeed, in the intermediate quantum regime, $g_2(a_1,b_1)\approx 1$. Computationally, it is difficult to map out the entire quantum-classical transition, but Fig.~\ref{fig:N2_g2} shows that $g_2(a_1,b_1)$ approaches 1 from the direction of its classical value. The presence of quantum noise and the lack of correlations are obvious in the quantum trajectories [Figs.~\ref{fig:N2_sync}(c) and \ref{fig:N2_sync}(d)].

However, Fig.~\ref{fig:N2_g2} shows that correlations reappear in the extreme quantum limit. This is because of entanglement between the cavity modes, which is a purely quantum effect. So although the classical correlations between the cavities are destroyed by quantum noise, they are replaced by quantum correlations. Note that $g_2(a_1,b_1)$ does not go directly from the classical value to the quantum value; rather, it first goes to 1, and then it goes to the quantum value. The quantum phase diagram is shown in Fig.~\ref{fig:N2_phasediagram}(b). The fact that the phase diagram is different from the classical result in Fig.~\ref{fig:N2_phasediagram}(a) demonstrates the importance of quantum correlations.

The physical reason for the correlations can be understood by examining the eigenstates of the coupled Hamiltonian in the absence of the laser drive: 
\begin{eqnarray}
H_2&=&i\frac{\chi}{2}(a_1^{\dagger2}a_2-a_1^2a_2^\dagger + b_1^{\dagger2}b_2-b_1^2b_2^\dagger)\nonumber\\
&&+V_1(a_1^\dagger b_1+a_1b_1^\dagger)+V_2(a_2^\dagger b_2+a_2b_2^\dagger). \label{eq:H2}
\end{eqnarray}
The eigenstates are shown in Fig.~\ref{fig:N1_levels}(b). The basis is $|m_1,m_2\rangle\otimes|n_1,n_2\rangle$, where $m_1,m_2,n_1,n_2$ are the Fock states for $a_1,a_2,b_1,b_2$, respectively. In the quantum limit, the only states populated are $|0,0\rangle\otimes|0,0\rangle$, $|1,0\rangle\otimes|0,0\rangle$, $|0,0\rangle\otimes|1,0\rangle$, and $|1,0\rangle\otimes|1,0\rangle$, since all other states have infinite level shifts due to $\chi$. The degeneracy of $|1,0\rangle\otimes|0,0\rangle$ and $|0,0\rangle\otimes|1,0\rangle$ is lifted by the coupling $V_1$. The new eigenstates are $|\pm\rangle=|1,0\rangle\otimes|0,0\rangle\pm|0,0\rangle\otimes|1,0\rangle$ with level shifts $\pm V$. Only $|+\rangle$ is coupled by the laser since the cavities are assumed to be identical. Whether $a_1$ and $b_1$ are bunched or antibunched depends on the relative population of $|1,0\rangle\otimes|1,0\rangle$. When $\Delta_1<0$, the laser is more resonant with $|+\rangle$ than with $|1,0\rangle\otimes|1,0\rangle$, leading to antibunching. When $\Delta_1>0$, the laser is more resonant with $|1,0\rangle\otimes|1,0\rangle$ than with $|+\rangle$, leading to bunching. This explains the structure of the quantum phase diagram in Fig.~\ref{fig:N2_phasediagram}(b).

Now we consider two nonidentical cavities ($\Delta_1^a\neq\Delta_1^b$). Given that two identical cavities have correlations in the quantum limit, one would expect the correlations to become weaker as the difference of $\Delta_1^a$ and $\Delta_1^b$ increases: after all, that is what happens in the classical case.
To find $g_2(a_1,b_1)$ in the quantum limit, we again solve Eq.~\eqref{eq:master} perturbatively in $E$ but now include $V_1$ and $V_2$. 
In the limit of $E\rightarrow 0$, $\chi\rightarrow\infty$, and small coupling,
\begin{eqnarray}
g_2(a_1,b_1)&=&1+\left[\frac{2\Delta_1^a}{\Delta_1^{a2}+\kappa_1^2} + \frac{2\Delta_1^b}{\Delta_1^{b2}+\kappa_1^2} \right.\nonumber \\
&& \quad\quad\quad - \frac{4(\Delta_1^a+\Delta_1^b)}{(\Delta_1^a+\Delta_1^b)^2+4\kappa_1^2}\bigg]V_1 +O(V_1^2).\quad \label{eq:g2a1b1_quantum}
\end{eqnarray}
This equation is plotted in Fig.~\ref{fig:N2_nonidentical}(b) as a function of the difference $\delta\equiv\Delta_1^a-\Delta_1^b$, keeping the average $\Delta_1\equiv(\Delta_1^a+\Delta_1^b)/2$ and the coupling fixed.

Equation \eqref{eq:g2a1b1_quantum} has three important features. 

(i) $V_1$ appears already in first order. This means that even for small coupling, there can still be significant correlations between the cavities. Note that unlike the classical case, there is no critical value of coupling, above which the correlations appear. This is because quantum noise smoothes out what was a sharp synchronization transition in the classical limit. 

(ii) When $\delta$ is small, $g_2(a_1,b_1)$ can be stronger than when $\delta=0$. (This occurs when $\Delta_1>\sqrt{3}\kappa_1$.) Thus, a slight mismatch of detunings can actually strengthen the correlation. This differs from the classical case, in which $g_2(a_1,b_1)$ monotonically approaches 1 for small $\delta$. 

(iii) When $\delta$ is large, $g_2(a_1,b_1)$ does not go to 1, but converges to $1-\frac{2\Delta_1V_1}{\kappa_1^2+\Delta_1^2}$. Thus, two very different cavities can still have correlations \footnote{When $\delta$ is so large that it is on the order of the drive frequency, the rotating-wave approximation breaks down, and one would have to use a time-dependent version of Eq.~\eqref{eq:H} that includes counter-rotating terms.}. This is quite surprising, since one would expect there to be no correlation when the cavities are far off resonance with each other. This classical intuition turns out to be wrong, because the cavities can be entangled despite the disparity in detuning. [Note that if one did not take the limit $\chi\rightarrow\infty$, $g_2(a_1,b_1)$ would go to 1 for large $\delta$.]

The quantum correlations arise because the laser excites entangled eigenstates of Eq.~\eqref{eq:H2}. When $\delta$ is large, the frequency $\Delta_1$ is still present in the system and is still near resonant with entangled eigenstates. Note that the physics behind the classical correlations is different; classical correlations exist when the frequency difference of the limit cycles is small relative to the coupling. Evidently, both quantum and classical correlations disappear in the intermediate regime when there is a lot of quantum noise.

\section{Conclusion}
We have shown that entanglement causes the cavities to behave quite unexpectedly in the quantum regime. Correlations exist between the cavities despite substantial quantum noise. Also, two very different cavities still have strong correlations, sometimes even stronger than when the cavities are identical. Our results suggest the possibility for a macroscopic number of oscillators to exhibit long-range correlations in the quantum limit. Classical synchronizing systems are known to exhibit phase transitions similar to equilibrium systems \cite{sakaguchi87,strogatz88,daido88,cross04,acebron05,hong05,lee09,lee10,chowdhury10,heinrich11}. One should investigate how quantum fluctuations affect these critical properties.

It would be interesting to study the quantum-classical transition for other quantum models known to exhibit limit cycles, as discussed in the introduction. However, our results are probably quite general, at least qualitatively. There should be classical correlations due to synchronization and quantum correlations due to entanglement, and the two regimes should be connected by an intermediate regime of almost no correlations.


We thank Mark Rudner and Sarang Gopalakrishnan for useful discussions. This work was supported by NSF through Grant No. DMR-1003337 and through a grant to ITAMP. 

\emph{Note added.} Recently we became aware of Refs.~\cite{ludwig12,mari13}, which study synchronization of optomechanics in the presence of quantum noise.

\bibliography{shg_arxiv}

\end{document}